# MoS$_2$ Dual-gate Transistors with Electrostatically Doped Contacts

Fuyou Liao[1, §], Yaocheng Sheng[1, §], Zhongxun Guo[1], Hongwei Tang[1], Yin Wang[1], Lingyi Zong[1], Xinyu Chen[1], Antoine Riaud[1], Jiahe Zhu[3], Yufeng Xie[1], Lin Chen[1], Hao Zhu[1], Qingqing Sun[1], Peng Zhou[1], Xiangwei Jiang[4], Jing Wan[2] (✉), Wenzhong Bao[1] (✉), David Wei Zhang[1]

[1] State Key Laboratory of ASIC and System, School of Microelectronics, Fudan University, Shanghai 200433, China
[2] State Key Laboratory of ASIC and System, School of Information Science and Engineering, Fudan University, Shanghai 200433, China
[3] School of Electronic Science and Engineering, Nanjing University, Nanjing 210093, China
[4] Institute of Semiconductors, Chinese Academy of Sciences, Beijing 100083 China
§ Fuyou Liao and Yaocheng Sheng contributed equally to this work

## ABSTRACT

Two-dimensional (2D) transition metal dichalcogenides (TMDs) such as molybdenum disulfide (MoS$_2$) have been intensively investigated because of their exclusive physical properties for advanced electronics and optoelectronics. In the present work, we study the MoS$_2$ transistor based on a novel tri-gate device architecture, with dual-gate (Dual-G) in the channel and the buried side-gate (Side-G) for the source/drain regions. All gates can be independently controlled without interference. For a MoS$_2$ sheet with a thickness of 3.6 nm, the Schottky barrier (SB) and non-overlapped channel region can be effectively tuned by electrostatically doping the source/drain regions with Side-G. Thus, the extrinsic resistance can be effectively lowered, and a boost of the ON-state current can be achieved. Meanwhile, the channel control remains efficient under the Dual-G mode, with an ON-OFF current ratio of 3×10$^7$ and subthreshold swing of 83 mV/decade. The corresponding band diagram is also discussed to illustrate the device operation mechanism. This novel device structure opens up a new way toward fabrication of high-performance devices based on 2D-TMDs.

## KEYWORDS

MoS$_2$, dual-gate, tri-gate, field effect transistor, extrinsic resistance, electrostatic doping

## 1 Instruction

In recent years, extensive research interest has been drawn into electrical and optical properties of two-dimensional (2D) materials, in particular, transition metal dichalcogenides (TMDs) [1-7]. Among them, semiconducting MoS$_2$ has been the most widely investigated, as it shows a thickness-dependent band gap with promising potential applications in transistors, memories, sensors, photodetectors, and light-emitters [8-15], which have been extensively investigated during the past few years. Recently, small and medium scale MoS$_2$ circuits have also been demonstrated, including logic gates, ring oscillator, 1-bit microprocessor and memory circuits [13, 16-18]. However, basic device processing of 2D-TMDs, such as doping of channel or contacts, is still far from practical application, which restricts the realization of large-scale circuits and more functionalities. A major obstacle is the metal contact/2D-TMDs interface engineering. For conventional bulk semiconductors, regular techniques such as precise control of ion implantation or diffusion and alloying of semiconductor with metals for Ohmic contacts have been investigated for decades and are broadly applied by the industry. However, these methodologies cannot be directly transferred to 2D-TMDs, mainly because they lack dangling bonds on the surface for a homogeneous reaction. Therefore, 2D-TMDs are extremely sensitive to any physical or chemical processing [19, 20], and it is challenging to form high-quality electrical contact and dielectric layer without affecting the intrinsic electrical properties.

Large contact resistance is currently a major issue to achieve high-performance TMDs based field effect transistors (FETs). The Schottky barrier (SB) at the metal-MoS$_2$ interface impedes the carrier injection efficiency from source to channel, thus it is a bottleneck for the flow of charge carriers and results in poor device performance [21-24]. Previous endeavors, such as the selection of metals with different work functions [21, 25], chemical doping of the source-drain [26-28], intercalation and oxide doping [29, 30], mechanical transfer of graphene or metal electrodes [31, 32] and Li-intercalation technique [33, 34], etc., have been confirmed effective for various TMDs materials. Nevertheless, most of these methods either demand complicated physical processing and chemical treatment, or show poor long-term stability. An alternative is via electrostatic doping by buried side gates (Side-G), which is relatively simple and has already been employed in FETs based on nanowire and carbon nanotube [35-43]. However, it has not yet been applied for 2D-TMDs based FETs.

Relatively low field-effect-mobility of TMDs also acts as the main bottleneck for device application. For example, high-speed device operation requires a large ON-state current, which is proportional to field-effect-mobility. For low mobility TMDs based FETs, a single back gate (BG) or top gate (TG) has limited tuning ability to improve the ON-state current, whereas the dual-gate (Dual-G) structure improves both ON-state current and electrostatic control in subthreshold region, similar to that of Fin-FET technology [44]. However, most of the previous studies focus on a Dual-G structure with a high-$k$ TG dielectric (usually $HfO_2$ and $Al_2O_3$) and a low-k global BG dielectric [45, 46]. The BG low-$k$ dielectric is often a 300 nm $SiO_2$ layer [47, 48], together with local TG they form an asymmetric Dual-G structure. Such asymmetric structure is rarely used for practical application, because under similarly applied voltages for both BG and TG, the channel is mainly controlled by the TG which has much larger gate capacitance. It is also difficult to integrate it in large-scale circuits as the BG is global and cannot be controlled separately for a single FET. Therefore, it is necessary to realize a symmetric or quasi-symmetric Dual-G structure for TMDs based high-performance FETs. By integration of the buried Side-G for source-drain and Dual-G for the channel, a tri-gate FET structure can be formed to combine both of their advantages.

Here we demonstrate a multilayer $MoS_2$ tri-gate FET with independent Dual-G that efficiently controls the channel and Side-G that tunes the SB together with non-overlapped regions. Under independent electrostatic doping from the Side-G on source-drain contacts, the SB together with extrinsic resistance $R_{ex}$ is fixed, and the electrical transport no longer follows the usual Schottky FET model applied for regular TMDs based FETs [49]. The electrical measurement of our device exhibits a large current density of 20.6 µA/µm, a low subthreshold swing (SS) of 83 mV/decade and an efficient modulation of the threshold voltage ($V_{TH}$) from -3 to 0.75 V. The ON/OFF current ratio is up to $3\times10^7$ for a thickness of 3.6 nm, even surpassing that of regular Schottky 2D-FETs.

## 2 Experimental section

The schematic diagram and electrical measurement connection of the $MoS_2$ tri-gate FET are shown in Fig. 1(a). Electron-beam (E-beam) lithography is used for all of the patterning steps and all metal deposition is via E-beam evaporation. The fabrication starts by patterning and metal deposition of Side-G electrodes (5 nm Ti and 30 nm Au) which locates at the bottom of the whole device. Atomic layer deposition (ALD) is then used to deposit a 15-nm-thick $HfO_2$ high-$k$ dielectric layer. Afterward, the BG electrode is formed and the gate oxide on top of BG is also a layer of 15-nm-thick $HfO_2$. Then, a multilayer $MoS_2$ sheet (thickness verified by atomic force microscope (AFM)) is mechanically exfoliated and transferred to the pre-patterned substrate, using a PDMS stamp and customer designed aligner (more details see Fig. S1 in the Electronic Supplementary Material). AFM image and height profile (~3.6 nm) of a transferred multilayer $MoS_2$ at this step are shown in Fig. 1(b) and (c). Next, Ti/Au (10 nm/40 nm) contact electrodes are patterned and deposited, followed by an immediate vacuum annealing at 200 ºC to remove the residual photoresist. A 2-nm-thick $Al_2O_3$ is then deposited as a seeding layer by E-beam evaporation, and afterward, a 15-nm-thick $HfO_2$ is grown by ALD as TG dielectric. Finally, a TG electrode (Au/40 nm) is formed with small gaps separating the TG and source/drain contacts. The schematic of the fabrication process flow of the tri-gate $MoS_2$ FET is shown in Fig. S2 in the Electronic Supplementary Material. An optical micrograph of an as-fabricated device is shown in Fig. 1(d). It is noteworthy that there parts contribute to the total resistance of the tri-gate $MoS_2$ FET, including contact resistance ($R_c$) from SB at the metal-$MoS_2$ interface, resistance from the non-overlapped region of channel ($R_{non-overlapped}$) and channel resistance ($R_{ch}$). Thus the extrinsic resistance ($R_{ex}$) is defined as $R_{ex}=R_c+R_{non-overlapped}$.

## 3 Results and discussions

In order to investigate the electrical characteristics, $MoS_2$ tri-gate devices are measured in the atmosphere using an Agilent B1500a parameter analyzer. We fabricated a batch of $MoS_2$ devices, all of which showed comparable electrical performance. The following data are collected from a device with a 3.6-nm-thick $MoS_2$ sheet. For this device, the basic output curves are measured first by sweeping the drain voltage $V_D$ from 0 to +2 V with a fixed $V_{Dual-G}$ =+4 V (here $V_{Dual-G} =V_{BG}=V_{TG}$), and $V_{Side-G}$ increases from -2 to +4 V with a step of 1 V. The results and an equivalent circuit model are shown in Fig. 2(a). Larger drain current $I_D$ and linear $I_D$-$V_D$ characteristics can be observed at large $V_{Side-G}$, which indicates that $V_{Side-G}$ can effectively tune the doping level of the source and drain regions. The $I_D$-$V_{Side-G}$ curve is also obtained by sweeping the Side-G from -4 to +4 V with a fixed $V_D$=0.1 V and $V_{BG}=V_{TG}$=0 V. Under this circumstance, the contact regions act as two transistors connected in series, and the $R_{ch}$ remains constant during the sweeping of $V_{Side-G}$. The current ON/OFF ratio ($I_{ON}/I_{OFF}$) is about $10^3$, which is also indicative of large tuning capability by $V_{Side-G}$.

Fig. 2(c) shows $I_D$ as a function of $V_D$ at various $V_{Dual-G}$ ranging from -2 to +4 V with $V_{Side-G}$=0 V. The non-linear $I_D$-$V_D$ characteristic at low bias and a faster saturation trend when increasing the $V_D$ suggests the current is limited by the large resistance of T1 and T2 (noted in the inset of Fig. 2(c)), especially the pinch-off effect of T2 [50]. The maximum $I_D$ is 15 µA and the corresponding current density is 1.5 µA/µm (channel length is 4 µm, channel width is 10 µm) under $V_D$=2 V and $V_{DG}$=+4 V. Fig. 2(d) displays $I_D$ as functions of $V_{BG}$, $V_{TG}$ and $V_{DG}$ when $V_{Side-G}$=0 V and $V_D$=0.1 V. The three curves exhibit a typical n-type behavior with a large $I_{ON}/I_{OFF}$ of ~$10^6$. The estimated SS under the BG-mode is about 120 mV/decade, and slightly larger for TG-mode due to the lower quality of the TG dielectric. The SS of Dual-G mode, however, is only 84 mV/decade, much smaller than single gate configuration, demonstrating a better electrostatic control. The two-probe field-effect mobility ($\mu_{2P}$) can be extracted from the linear region of the transfer curves using the following equation [51]:

$$\mu_{2P} = \frac{dI_D}{dV_G} \times \frac{L}{WC_{ox}V_D} \quad (1)$$

where $L$ and $W$ are the length and width of the channel, respectively, $C_{ox}$ is the gate dielectric capacitance per unit area (the gate dielectric for Dual-G mode is about $2C_{ox}$). The $\mu_{2P}$ for the BG, TG and Dual-G mode is estimated as 4, 3 and 3.5 cm$^2$/(V·s), respectively. Next, the output $I_D$-$V_D$ curves under Dual-G mode with $V_{Side-G}$=+4 V are shown in Fig. 2(e). The near-ohmic behavior and significantly larger $I_D$ are in stark contrast with the measurements under $V_{Side-G}$=0 V (Fig. 2(d)). This suggests that increasing $V_{Side-G}$ narrows the SB width and/or decreases its effective height, but also lowers the resistance of the non-overlapping region. Under high $V_{Dual-G}$, the saturation of $I_D$ in Fig. 2(c) with $V_{Side-G}$=0 V is more pronounced than that in Fig. 2(e) under $V_{Side-G}$=+4 V. This is due to that the transistor T2 controlled by $V_{Side-G}$ reaches saturation early with a low $V_{Side-G}$. Whereas the T2 operates in a linear region with a high $V_{Side-G}$, and thus the saturation of the transistor is mainly determined by the transistor T3 which is controlled by $V_{Dual-G}$.

Fig. 2(f) shows $I_D$ as functions of $V_{BG}$, $V_{TG}$ and $V_{Dual-G}$, with $V_{Side-G}$=+4 V. Better switching properties with a high $I_{ON}/I_{OFF}$ of ~$10^7$ are achieved, which is much larger than previously reported results for MoS$_2$ with the same thickness [52, 53]. The SS at $V_D$=0.1 V for the BG and Dual-G mode is 110 and 83 mV/decade, respectively, which are comparable to those of previously reported values [53]. The two-terminal $\mu_{2P}$ of the BG, TG and Dual-G mode under $V_{Side-G}$=+4 V is estimated to be 8.7, 7.3 and 8.5 cm$^2$/(V·s), respectively, which is more than twice of the mobility when $V_{Side-G}$=0 V. This is mainly due to the improvement of $R_{ex}$, which makes the extraction of mobility more accurate (to be discussed below). We also notice that as expected, under the Dual-G mode the ON-state $I_D$ is much larger than that of the single gate mode. It indicates that the nearly identical curves shown in Fig. 2(d) can be explained by the domination of the T1 and T2 under large resistance. The above results show that the tri-gate MoS$_2$ FET has an improved electrical performance with $V_{Side-G}$=+4 V compared to that when $V_{Side-G}$=0 V. This improvement tuned by $V_{Side-G}$ stems from (i) the high doping of the non-overlapped region and (ii) the reduction of the SB between MoS$_2$ and metal contacts. Thus, the Side-G can work as a tuning knob to switch the device between the low power consumption (by increasing $R_{ex}$) and the high-speed mode (by decreasing $R_{ex}$).

To further explore electrical properties our tri-gate MoS$_2$ FET, transfer curves under Dual-G mode with varying $V_{Side-G}$ are characterized. As shown in Fig. 3(a), when $V_{Side-G}$ increases from 0 to +4 V, $I_{ON}$ is boosted from 2.3 to 15 μA, while the $V_{TH}$ remains almost the same. Fig. 3(b) is the corresponding semi-log scaled $I_D$-$V_{Dual-G}$ curves. It noteworthy that while the $I_{OFF}$ remains almost the same, the increase of $I_{ON}$ contributes to the improved $I_{ON}/I_{OFF}$. Extracted values of $I_{ON}/I_{OFF}$ and SS as a function of $V_{Side-G}$ are shown in Fig. 3(c). $I_{ON}/I_{OFF}$ has a nearly linear dependence with $V_{Side-G}$, and the SS scatters between 83~90 mV/decade, which is almost independent of $V_{Side-G}$. To quantitatively investigate the mobility and $R_{ex}$ as a function of $V_{Side-G}$, intrinsic mobility ($\mu_0$) and $R_{ex}$ can be extracted using either rational function fitting or Y-function method [54, 55] (see Electronic Supplementary Material Fig. S7-8). As shown in Fig. 3(d), $\mu_{2P}$ increases from 3.5 to 8.5 cm$^2$/(V·s) due to the effect of $R_{ex}$, $\mu_0$≈10 cm$^2$/(V·s) extracted by rational function fitting method is independent of $V_{Side-G}$, and $R_{ex}$ decreases from 190 to 14 kΩ·μm, which again suggests that normally extracted $\mu_{2P}$ is significantly underestimated due to large $R_{ex}$ (comparable to the $R_{ch}$). The electrical properties of tri-gate devices with varied MoS$_2$ thickness are also measured, details can be found in Electronic Supplementary Material Fig. S4-6.

Based on the above experimental results, schematics band diagrams can be proposed to illustrate the doping effect from Side-G. As shown in Fig. 3 (e-f), the electrons injected from the source electrode have to overcome the SB at the metal-MoS$_2$ interface. When applying a lower $V_{Side-G}$ the depletion region at the metal-MoS$_2$ interface grows larger, leading to a smaller drain current, which agrees well with the results shown in Fig. 3(e). When a larger positive $V_{Side-G}$ is applied, the pulled down energy band reduces the effective SB height. Thus, electrons are easier to inject from the source electrode to the channel, which gives rise to a larger $I_D$.

We then investigate the gate modulation mechanism in the Dual-G architecture. While maintaining a constant $V_{Side-G}$ of +4 V, $I_D$ is mapped by varying both $V_{BG}$ and $V_{TG}$, as shown in Fig. 4(a), respectively. Within the region of $V_{TG}$<1 V and $V_{BG}$>-1.5 V, the tangent slope (≈1) indicates that controllability is approximately the same for $V_{TG}$ and $V_{BG}$. This also agrees well with the estimated capacitance ratio of BG to TG ($C_{BG}/C_{TG}$≈1) based on the capacitance measurement. Above $V_{TG}$=1 V and $V_{BG}$<-1.5 V (upper left region), the tangent becomes nearly vertical, indicating that $I_D$ is now mainly modulated by the BG. This abnormal phenomenon is attributed to the imperfect symmetric Dual-G device structure, which can be more clearly shown in Fig. 4(b). Since the thickness of the MoS$_2$ sheet is only 3.6 nm and the BG electrode is 35-nm-thick, after the MoS$_2$ sheet is transferred onto the pre-patterned substrate with BG, the MoS$_2$ follows its surface morphology and bends at the edge of the BG electrode. Therefore, the edge of the BG is able to modulate the vertical fragment of the MoS$_2$, but the gating efficiency from the TG is drastically reduced. Once the BG turns off the current in MoS$_2$, the TG can modulate the horizontal MoS$_2$ channel but has little influence on the vertical fragment of MoS$_2$, so $I_D$ is now dominated by BG.

A more detailed plot of transfer characteristics extracted from Fig. 4(a) is displayed in Fig. 4(c). These characteristics are obtained by sweeping $V_{BG}$ while fixing the $V_{TG}$ at a series of constant values. The linear shift of the BG $V_{TH}$ as a function of $V_{TG}$ is shown in Fig. 4(d). The value of $V_{TH}$ is extracted from Fig. 4(c) following equation $V_{TH}$=$V'_{TH}$-(1/2)$V_D$ , where $V'_{TH}$ is the intercept with the X-axis by fitting linear region [56]. The Fig. 4(d) shows that the $V_{TH}$ of BG can be effectively modulated from 0.75 to -3 V by tuning $V_{TG}$ from -4 to 1 V. This characteristic is also attractive for switching the circuits between high performance and low power consumption modes [57]. With increasing $V_{TG}$ from -4 to +4 V, the $\mu_{2P}$ increases from 5 to 14 cm$^2$/(V·s), as shown in Fig. 4(d). The increase of $V_{TG}$ leads to higher electron concentration, which can partially screen the charge impurities and trapped charges at the MoS$_2$-HfO$_2$ interface, and thus results in higher mobility [58].

# 4 Conclusion

We demonstrate the fabrication of a multilayer MoS$_2$ tri-gate FET. This device exhibits high performances and attractive features, such as low $R_{ex}$, low operating gate voltage and tunable $V_{TH}$. On the one hand, the $R_{ex}$ can be tuned from 190 to 14 kΩ·μm by applying a positive $V_{Side-G}$, thanks to electrostatic doping near the source/drain regions. On the other hand, the Dual-G mode has excellent electrostatic control on the MoS$_2$ channel ($I_{ON}/I_{OFF}$=3×10$^7$, SS=83 mV/decade, $\mu_{2P}$=14 cm$^2$/(V·s)) than the single BG and TG mode. The operation of this device has been explained in details by the band diagrams. Same operation mechanism is expected to apply to other 2D semiconductors and may contribute to low-power and high-speed devices based on 2D materials in the future.

# Acknowledgements

This work was supported by the National Key Research and Development Program of China (2016YFA0203900, 2018YFA0306101), Shanghai Municipal Science and Technology Commission (18JC1410300) and Natural Science Foundation of China (Grant No: 61874154).

**Electronic Supplementary Material**: Supplementary material (photograph of the customer designed aligner, details manufacturing process of the devices, the resistance model, the electrical properties of tri-gate devices with varied MoS$_2$ thickness, rational function fitting method and Y-function method) is available in the online version of this article at http://dx.doi.org/10.1007/s12274-***-****-* (automatically inserted by the publisher).

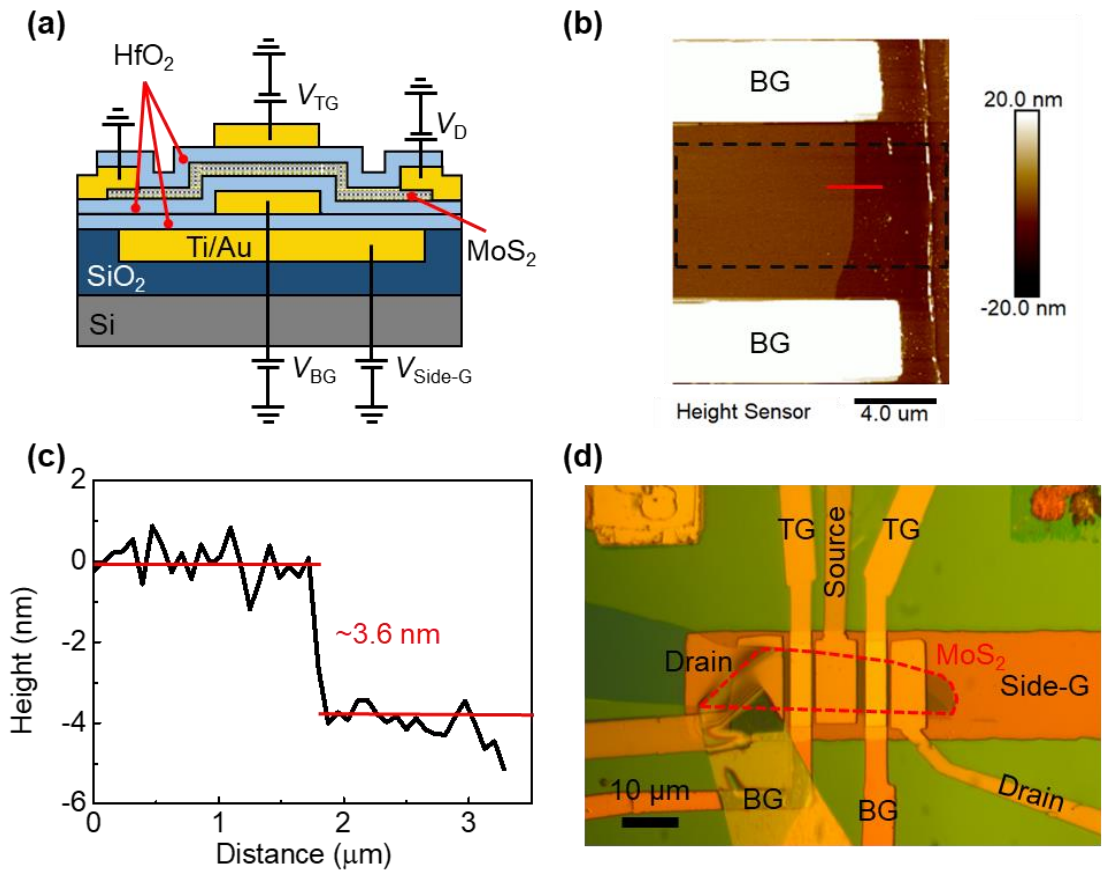

**Figure 1** (a) Schematics of the MoS$_2$ tri-gate transistor with electrical connections used to characterize the device. (b) AFM image of the MoS$_2$ flake after transferring to the buried BG electrodes. The region indicated by a black dashed box will be filled by depositing contact electrode during the next fabrication step. (c) Height profile of the multilayer MoS$_2$. The height profile is measured along the red line in (b). (d) Optical image of the fabricated device (two FETs connected in series based on the same MoS$_2$ flake).

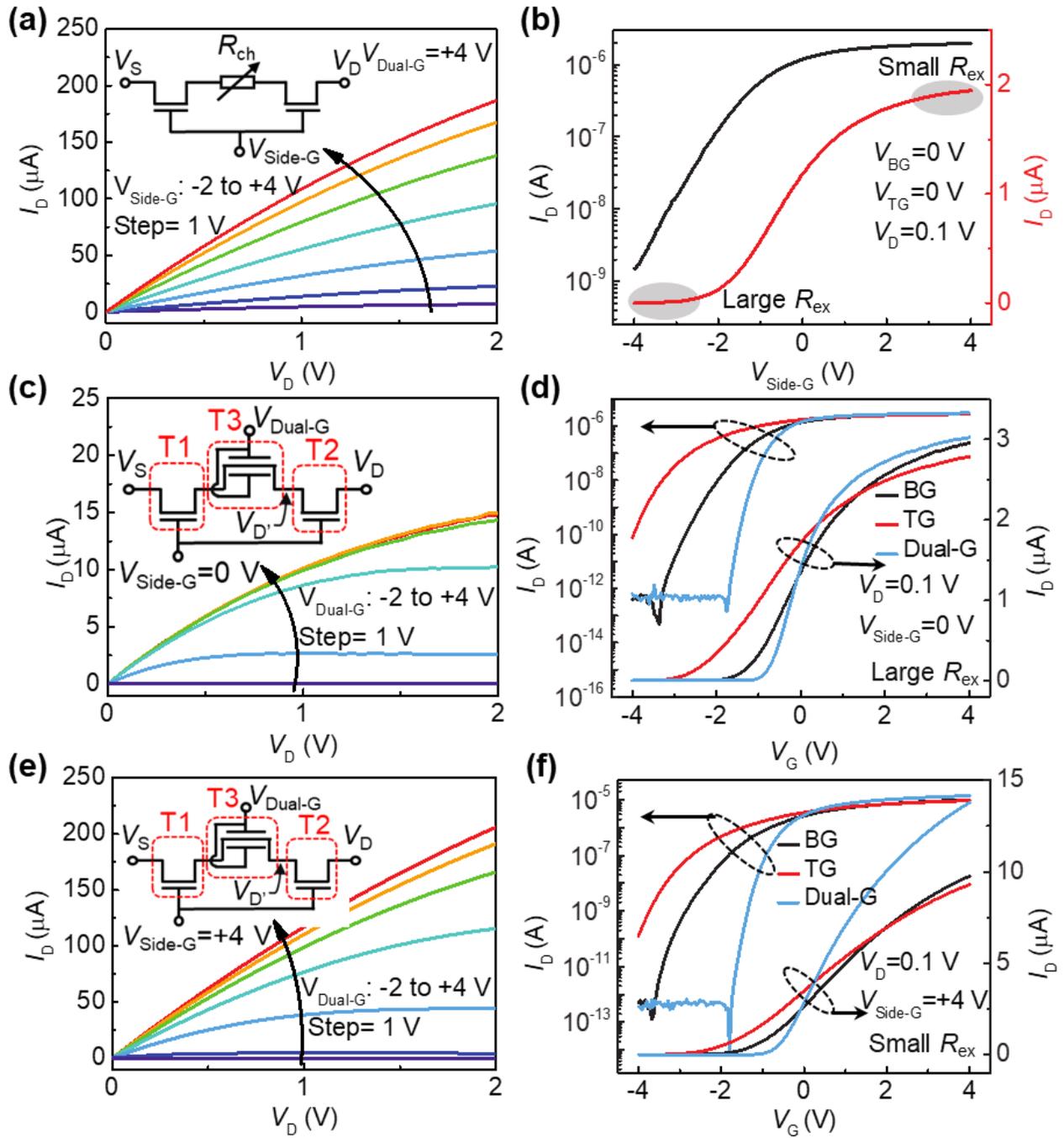

**Figure 2** (a) Output characteristic curves of MoS$_2$ Side-G FET with $V_{Dual-G}$=+4 V. The inset shows an equivalent circuit model of the MoS$_2$ tri-gate transistor. (b) Transfer characteristic curves of MoS$_2$ Side-G FET obtained by sweeping the $V_{Side-G}$ from -4 to +4 V with a fixed bias of $V_D$=0.1 V and $V_{BG}$=$V_{TG}$=0 V. (c) Output characteristic curves of MoS$_2$ tri-gate FET for various applied $V_{Dual-G}$ with $V_{Side-G}$=0 V. The inset shows an equivalent circuit model of the MoS$_2$ tri-gate transistor with a low $V_{SG}$. (d) Transfer characteristic curves of $V_{BG}$, $V_{TG}$ and $V_{Dual-G}$ with $V_D$=0.1 V and $V_{Side-G}$=0 V at room temperature. (e) Output characteristic curves of MoS$_2$ tri-gate FET for various applied $V_{Dual-G}$ with $V_{Side-G}$=+4 V. The inset shows an equivalent circuit model of the MoS$_2$ tri-gate transistor with a high $V_{Side-G}$. (f) Transfer characteristic curves of $V_{BG}$, $V_{TG}$ and $V_{Dual-G}$ with $V_D$=0.1 V and $V_{Side-G}$=+4 V at room temperature.

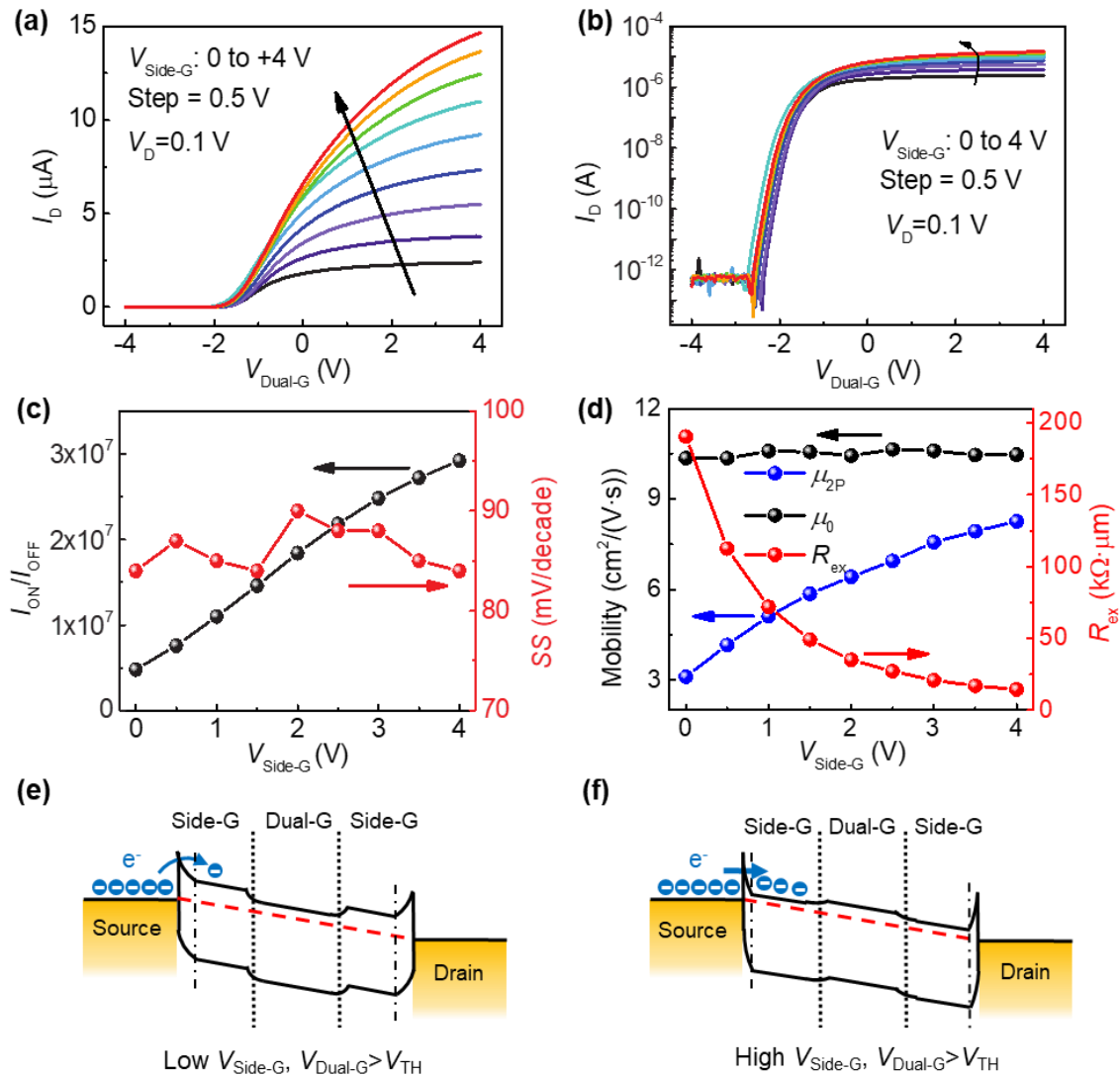

**Figure 3** $V_{Side-G}$ dependent electrical measurement. Linear (a) and semi-log (b) plot of $I_D$-$V_{DG}$ characteristic of the MoS$_2$ tri-gate transistor under various applied $V_{Side-G}$ from 0 to +4 V at a step of 0.5 V. (c) $I_{ON}/I_{OFF}$ and extracted values of $SS$ as a function of $V_{Side-G}$. (d) Extracted values of $\mu_{2P}$, $\mu_0$ and $R_{ex}$ as a function of $V_{Side-G}$. Band-diagram of the device when applying low $V_{Side-G}$ (e) and high $V_{Side-G}$ (f) with $V_{Dual-G}$>$V_{TH}$.

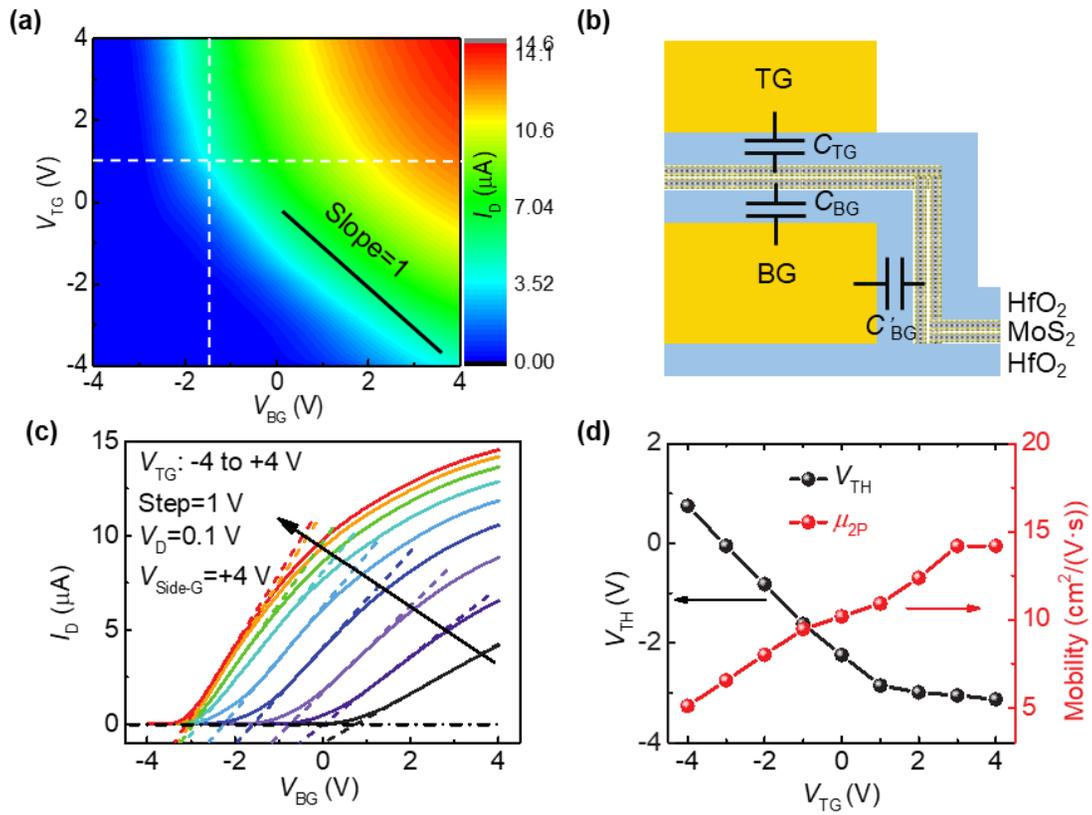

**Figure 4** The electrical characteristic of MoS$_2$ tri-gate FET with $V_{Side-G}$=+4 V. (a) 2D contour plot of $I_D$ as a function of $V_{TG}$ and $V_{BG}$ at constant drain voltage $V_D$=0.1 V. (b) A simplified capacitance model of the tri-gate FETs. (c) Linear plots of BG transfer characteristics with TG voltage ranging from -4 to +4 V. The step of gate voltage change is 1 V. The dashed lines indicate changes in the slope of d$I_D$/d$V_{TG}$. (d) $V_{TH}$ and $\mu_{2P}$ (extracted from BG transfer curves (c)) as a function of TG bias.

**Electronic Supplementary Material**

# MoS$_2$ Dual-gate Transistors with Electrostatically Doped Contacts

Fuyou Liao[1, §], Yaocheng Sheng[1, §], Zhongxun Guo[1], Hongwei Tang[1], Yin Wang[1], Lingyi Zong[1], Xinyu Chen[1], Antoine Riaud[1], Jiahe Zhu[3], Yufeng Xie[1], Lin Chen[1], Hao Zhu[1], Qingqing Sun[1], Peng Zhou[1], Xiangwei Jiang[4], Jing Wan[2] (✉), Wenzhong Bao[1] (✉), David Wei Zhang[1]

[1] *State Key Laboratory of ASIC and System, School of Microelectronics, Fudan University, Shanghai 200433, China*
[2] *State Key Laboratory of ASIC and System, School of Information Science and Engineering, Fudan University, Shanghai 200433, China*
[3] *School of Electronic Science and Engineering, Nanjing University, Nanjing 210093, China*
[4] *Institute of Semiconductors, Chinese Academy of Sciences, Beijing 100083 China*
[§] *Fuyou Liao and Yaocheng Sheng contributed equally to this work*

**Section 1**

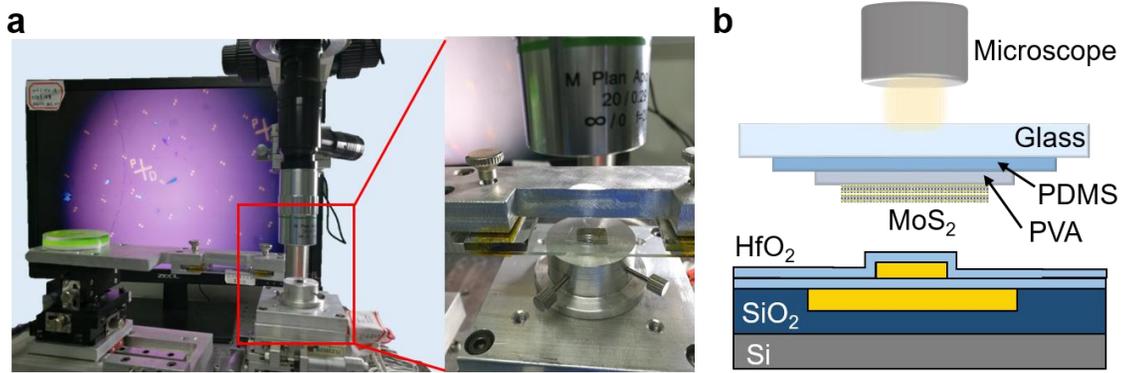

**Figure S1** Photograph (a) and schematic diagram (b) of the customer designed micro-aligner.

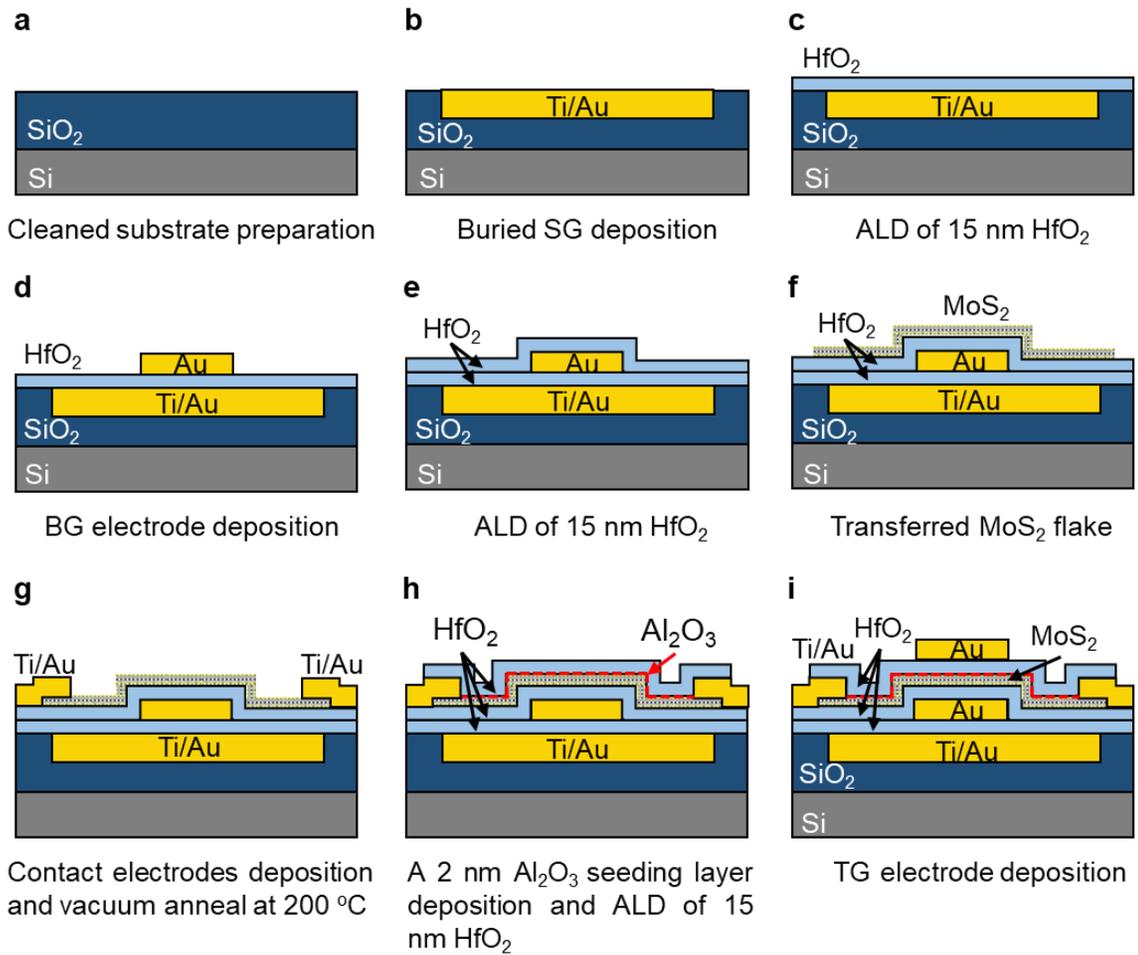

**Figure S2** (a-i) Fabrication process flow of the tri-gated $MoS_2$ FET.

Address correspondence to Jing Wan, email: jingwan@fudan.edu.cn ; Wenzhong Bao, email: baowz@fudan.edu.cn

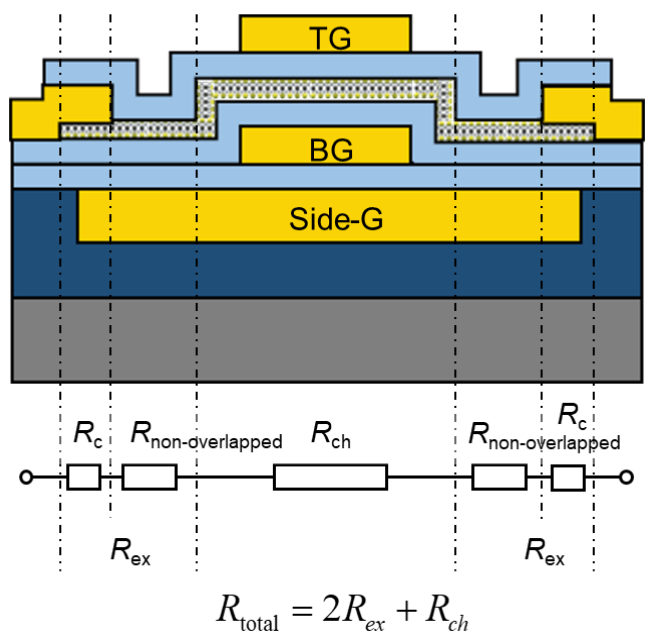

$$R_{\text{total}} = 2R_{ex} + R_{ch}$$

**Figure S3** A schematic cross sectional view the tri-gate MoS$_2$-FET (upper) and the corresponding resistance model (lower).

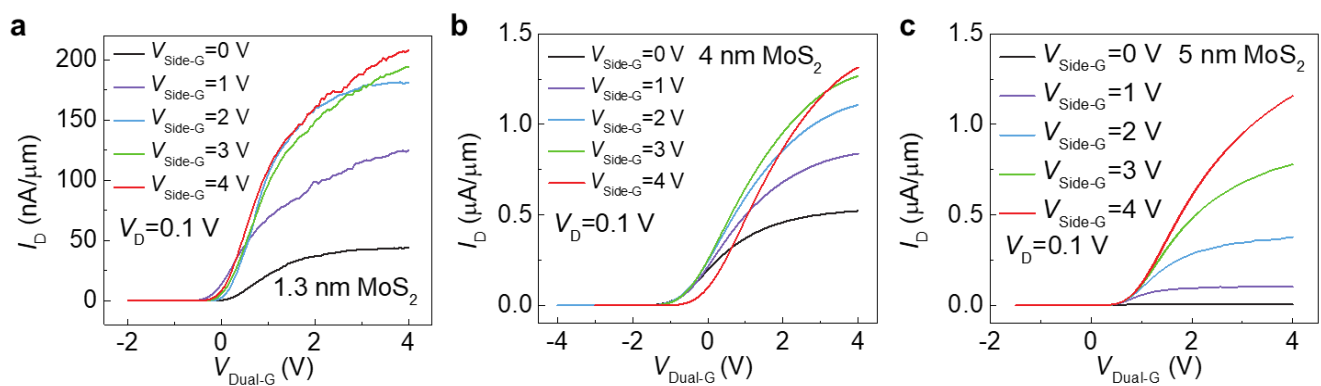

**Figure S4** Transfer characteristics of MoS$_2$ tri-gate FETs with varied $V_{\text{Side-G}}$ values. The thickness of the MoS$_2$ channel for each panel is 1.3 nm (a), 4 nm (b) and 5 nm (c).

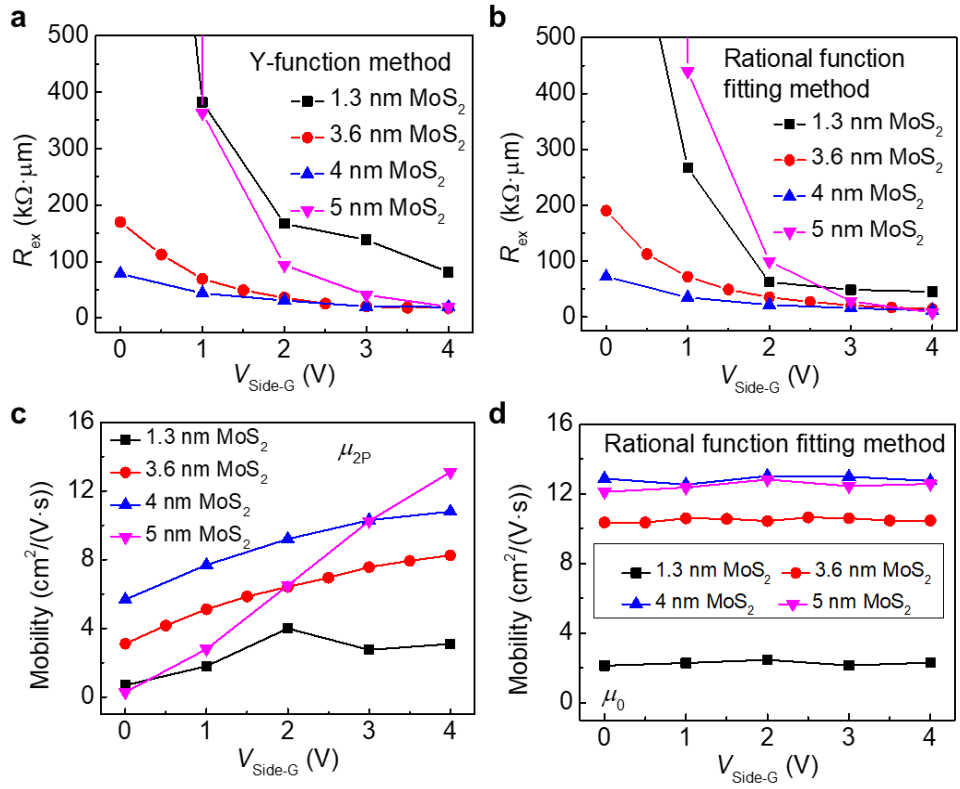

**Figure S5** Thickness dependence of $R_{ex}$, $\mu_{2P}$ and $\mu_0$. (a-b) Calculated $R_{ex}$ of the tri-gate MoS$_2$ FETs as a function of $V_{Side-G}$. The extraction of $R_{ex}$ is based on a Y-function method (a) and rational function fitting method (b). (c) $\mu_{2P}$ as a function of $V_{Side-G}$. (d) Intrinsic mobility $\mu_0$ of the devices extracted by rational function fitting method.

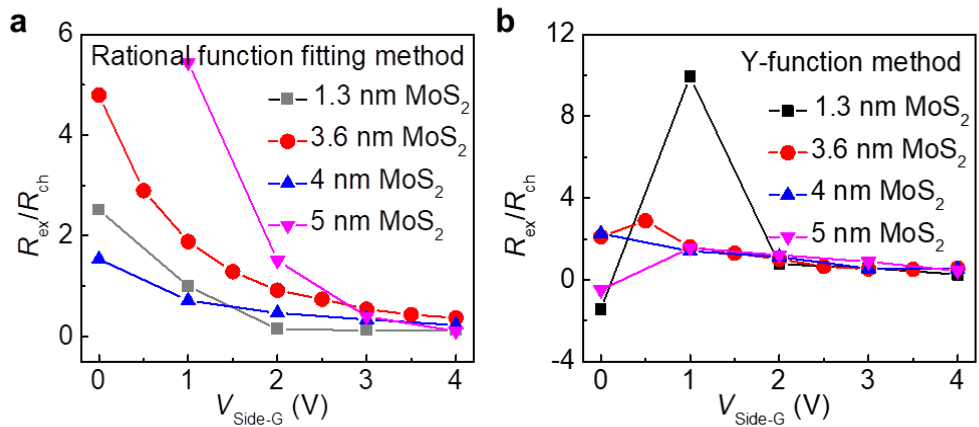

**Figure S6** Ratio of $R_{ex}/R_{ch}$ as a function of $V_{Side-G}$. $R_{ex}$ is extracted by rational function fitting (a) and the Y-function method (b). The results indicate that the $R_{ex}$ extracted by the rational fitting method is more reasonable than that of the Y-function method.

## Section 2
## Rational function fitting method

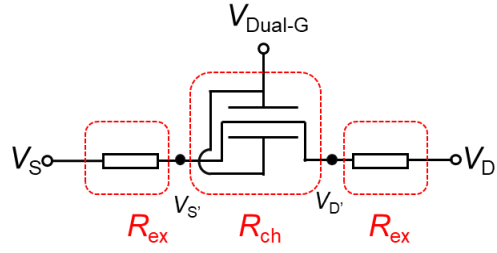

**Figure S7** Schematic of equivalent circuit model of the tri-gate MoS$_2$ FET with a fixed $V_{\text{Side-G}}$.

As the simplified equivalent circuit model shown in Fig. S7,

$$R_{total} = R_{ch} + 2R_{ex} = \frac{V_D}{I_D} \quad (S1)$$

where $R_{total}$ is the total resistance and $R_{ch}$ is channel resistance, $R_{ex}$ is extrinsic resistance relate to $V_{SG}$, including the resistance from the ungated regions ($R_{\text{non-overlapped}}$) and contact resistance ($R_c$) between metal and MoS$_2$, $V_D$ is drain voltage bias and $I_D$ is the drain current. $I_D$ can also be written as:

$$I_D = 2\mu_0 C_{ox} \frac{W}{L}(V_{GS'} - V_{TH})V_{D'} \quad (S2)$$

Where $\mu_0$ is intrinsic carrier mobility, $2C_{ox}$ is the dual-gate oxide capacitance, $W$ and $L$ is the channel width and length, $V_{D'}$ is effective drain bias of the T3, denoted in Fig. S7. For small $V_D$, the $V_{GS'}$ is approximately equal to V$_{GS}$, thus

$$R_{ch} = \frac{V_{D'}}{I_D} = \frac{1}{2\mu_0 C_{ox}\frac{W}{L}(V_{GS} - V_{TH})} \quad (S3)$$

$$R_{total} = \frac{\frac{L}{2\mu_0 C_{ox} W}}{V_{GS} - V_{TH}} + 2R_{ex} \quad (S4)$$

Therefore, it can be simplified as a rational function:

$$y = \frac{a}{x-b} + c \quad (S5)$$

where $a = \frac{L}{2\mu_0 C_{ox} W}$, $b = V_{TH}$ and $c = 2R_{ex}$. We can use equation (S5) to fit the $R_{total} \sim V_{GS}$ curve and obtain the fitting parameter $c$.

## Section 3
## Y-function method [1-6]

The Y-function is defined as

$$Y = \frac{I_D}{\sqrt{g_m}} \tag{S6}$$

$I_D$ is the drain current, and $g_m$ is the transconductance ($= \frac{dI_D}{dV_G}$), Therefore, the Y-function can be calculated from the transfer characteristics ($I_D$-$V_G$). In the strong inversion region, the Y-function is linearly dependent on $V_G$ as:

$$Y = \frac{I_D}{\sqrt{g_m}} = \left(2\mu_0 C_{ox} V_D \frac{W}{L}\right)^{0.5} (V_G - V_{TH}) \tag{S7}$$

where $\mu_0$ is the low field mobility, $2C_{ox}$ is the gate oxide capacitance, $V_D$ is the drain voltage, and $V_{TH}$ is the threshold voltage. $W$ and $L$ are the channel width and length, respectively. After the linear fitting of the Y-function vs. $V_G$, both low field mobility ($\mu_0$) and threshold voltage ($V_{TH}$) can been extracted. Once $\mu_0$ and $V_{TH}$ are obtained, the mobility attenuation coefficient ($\theta$) can be determined as:

$$\theta = \left[\frac{I_D}{g_m (V_G - V_{TH})} - 1\right] \bigg/ (V_G - V_{TH}) \tag{S8}$$

The mobility degradation factor ($\theta$) includes the effects of the extrinsic resistance ($2R_{ex}$), and can be expressed as:

$$\theta = \theta_0 + 4 R_{ex} \mu_0 C_{ox} \frac{W}{L} \tag{S9}$$

where $\theta_0$ is the intrinsic mobility degradation factor. The $\theta$ in multilayer MoS$_2$ has a significantly larger contribution from $2R_{ex}$ as compared to the $\theta_0$. Hence, $\theta_0$ is considered negligible [2, 3]. As a result, the contact resistance $R_{ex}$ can be calculated from $\theta$.

$$R_{ex} = \frac{\theta L}{4 W \mu_0 C_{ox}} \tag{S10}$$

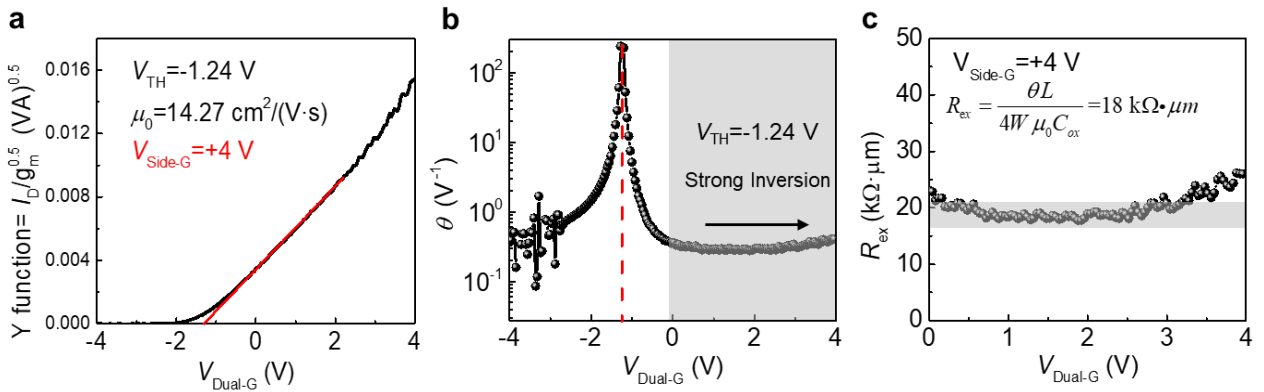

**Figure S8** Extraction of $R_{ex}$ by the Y-function method. (a) Y-function vs. $V_{\text{Dual-G}}$ of the tri-gate MoS$_2$ transistor with Dual-G mode and $V_{\text{Side-G}}$=+4 V. From the linear fit in the strong inversion region (red solid line), both the low-field mobility ($\mu_0$) and threshold voltage ($V_{TH}$) can be extracted from the x-intercept and the slope. (b) The mobility attenuation coefficient ($\theta$) as a function of $V_{\text{Dual-G}}$ can be calculated by the equation (3). (c) The $R_{ex}$ as a function of $V_{\text{Dual-G}}$ which can be calculated from $\theta$.